\newif\ifconfver
\confverfalse      

\newif\ifcutshort      
\cutshorttrue

\newif\ifcutshortlvltwo  
\cutshortlvltwofalse

\ifconfver
    \documentclass[10pt,twocolumn,twoside]{IEEEtran}
\else
    \documentclass{IEEEtran}
\fi

\usepackage{cite}
\usepackage{graphicx}
\usepackage{psfrag}
\usepackage{url}
\usepackage{stfloats}
\usepackage{amsfonts,amssymb,amsmath,bm,paralist,theorem,cite,ifthen,color}
\usepackage{array}
\usepackage{caption}
\usepackage{calc}
\usepackage{subfigure}
\usepackage{longtable}
\usepackage{stfloats}  
\usepackage{graphics,booktabs,color,epsfig,subfigure,enumerate}
\usepackage{multirow}
\usepackage{algorithm,algorithmic}
\usepackage{epstopdf}
\usepackage{array}
\usepackage{stfloats}
\usepackage{diagbox}

\theorembodyfont{\rmfamily}

{\begin{list}{}{
    \settowidth{\labelwidth}{\mbox{\textnormal{#1}}}%
    \setlength{\leftmargin}{\labelwidth+\labelsep}}}%
{\end{list}}

\begin{document}

\bibliographystyle{IEEEtran}

\newcommand\bcc[2][c]{\ensuremath{\bm{\mathcal{#2}}}}      
\newcommand\bcl[2][c]{\ensuremath{\bm{#2}}}
\newcommand\Ib{\ensuremath{{\bm I}}}
\newcommand\Vb{\ensuremath{{\bm V}}}
\newcommand\vb{\ensuremath{{\bm v}}}
\newcommand\Hb{\ensuremath{{\bm H}}}
\newcommand\ub{\ensuremath{{\bm u}}}
\newcommand\hb{\ensuremath{{\bm h}}}
\newcommand\xb{\ensuremath{{\bm x}}}
\newcommand\gb{\ensuremath{{\bm g}}}
\newcommand\Thetab{\ensuremath{{\bm \Theta}}}
\newcommand\thetab{\ensuremath{{\boldsymbol \theta}}}
\newcommand\Gb{\ensuremath{{\bm G}}}
\newcommand\Pb{\ensuremath{{\bm P}}}
\newcommand\taub{\ensuremath{{\boldsymbol \tau}}}
\newcommand\zb{\ensuremath{{\bm z}}}
\newcommand\st{\ensuremath{{\rm ~s.t.}}}


\def\blue{\color{blue}}
\def\red{\color{red}}
\definecolor{orange}{RGB}{255,107,0}
\def\orange{\color{orange}}

\title{One-Bit Symbol-Level Precoding for MU-MISO Downlink with Intelligent Reflecting Surface}

\ifconfver \else {\linespread{1.1} \rm \fi

\author{ Silei Wang, Qiang Li and Mingjie Shao
	\thanks{S. Wang and Q. Li are with
		School of Information and Communication  Engineering, University of Electronic Science and Technology of China, Chengdu, P.~R.~China, 611731. E-mail: wangniuqing520@gmail.com, lq@uestc.edu.cn ( {\it Corresponding author: Qiang Li}).}
		\thanks{M. Shao is with Dept. Electronic Engineering, The Chinese University of Hong Kong, Hong Kong. E-mail: mjshao@ee.cuhk.edu.hk}
}
\maketitle

\begin{abstract}

This paper considers  symbol-level precoding (SLP) for multiuser multi-input single-output (MISO) downlink transmission with the aid of intelligent reflecting surface (IRS). Specifically, by assuming one-bit transmitted signals at the base station (BS), which arises from the use of low-resolution DACs in the regime of massive transmit antennas, a joint design of one-bit SLP at the BS and the phase shifts at the IRS is proposed with a goal of minimizing the worst-case symbol error probability (SEP) of the users under the PSK modulation. This joint design problem is essentially a mixed integer nonlinear program (MINLP). To tackle it, we alternately optimize the one-bit signal and the phase shifts. For the former, a dual of the relaxed one-bit SLP problem is solved by  the mirror descent (MD) method with the maximum block improvement (MBI) heuristics. For the latter, the  accelerated projected gradient (APG) method is employed to optimize the phases. Numerical results demonstrate  that the proposed joint design can attain better SEP performance than  the conventional linear precoding and one-bit SLP.

%
%
%
%
%
%
%
\end{abstract}


\ifconfver \else \IEEEpeerreviewmaketitle} \fi

\section{Introduction}
The network capacity proliferation and ubiquitous wireless connectivity in the forthcoming wireless network pose great challenges to the existing wireless transmission technology. Seeking for new techniques with low cost, high spectral and energy efficiency is crucial for 5G and 6G networks~\cite{wu2017overview}. To this end, intelligent reflecting surface (IRS) has recently been brought forward as a promising cost-and-energy efficient solution. IRS is a software-controlled metasurface (a planar array) composed of numerous low-cost passive elements, each of which is embedded several electronic devices, such as positive-intrinsicnegative (PIN) diodes, field-effect transistors (FETs). Triggered by a smart controller integrated in the IRS, each element is able to independently reflect the incident electromagnetic wave with preprogrammed phase shifts and/or reflecting amplitudes such that the reflected signal can be beneficially (resp. destructively) aligned at the intended (resp. non-intended) receiver, which is referred to as software-controlled smart radio environments~\cite{wu2019towards, liaskos2018new}.

There have been a flow of works on IRS. In~\cite{wu2019intelligent,guo2020weighted}, the authors considered incorporating IRS into the standard multiuser downlink transmission to either reduce the transmit power at the base station (BS)~\cite{wu2019intelligent} or increase the system sum rate~\cite{guo2020weighted}. Both works require jointly optimizing the transmit beamformers at the BS and the phase shifts at the IRS.
More recently, IRS has also been studied in various wireless related applications, such as physical-layer security~\cite{chen2019intelligent,guan2020intelligent}, non-orthogonal multiple access (NOMA) systems~\cite{zuo2020resource}, simultaneous wireless information and power transfer~\cite{wu2020joint}, to name a few.

It is worthwhile to mention that most of the existing works on IRS focus on linear precoding scheme. However, conventional linear precoder is based on fully-digital signal processing with high-resolution digital-to-analog converters (DACs), which could induce prohibitive hardware cost and power consumption in the case of large-scale antenna arrays, e.g., massive multiple-input multiple-output (MIMO). As such, using low-resolution DACs, especially one-bit DACs, in massive MIMO is seen as a promising way to circumvent this problem. Under the one-bit signaling, the design of the transmitted signal is performed in  a per-symbol time manner by utilizing both the symbol information and channel state information (CSI) to directly synthesize the desired symbols at the receivers.
%
Symbol error probability (SEP) and mean squared error (MSE) are usually adopted as the design metrics for one-bit symbol-level precoding (SLP). Interested readers are referred to~\cite{shao2019framework,jacobsson2017quantized} for more details.

One-bit SLP for massive MIMO uplink/downlink has been widely studied and proved to achieve prominent performance~\cite{choi2016near,shao2018one,shao2018multiuser}. However, as of this writing we are not aware of any work concerning the  one-bit SLP with IRS. This work is one of the earliest attempts to study the effectiveness of the IRS in one-bit SLP. The most relevant  works are those in~\cite{liu2019joint, shao2020minimum}, where  SLP with IRS  is studied without considering the one-bit constraint. In light of this, this paper considers an IRS-assisted multiuser multiple-input single-output (MISO) downlink communications, where one-bit DACs are deployed at the BS.
We jointly design the one-bit transmitted signals at the BS and the phase shifts at the IRS, so that  the worst-case SEP of all the users is minimized under the PSK modulation. The resulting problem is non-convex and intractable. To tackle this min-max SEP problem, we first transform the SEP function into a more tractable form via upper bound approximation~\cite{shao2018multiuser}. Then, we alternately optimize the one-bit signals and the phase shifts. Specifically, for the one-bit transmitted signals design, we first relax the one-bit constraint and transform the relaxed problem into its dual form, which can be efficiently solved via the mirror descent (MD) method~\cite{beck2003mirror}. After that, the maximum block improvement (MBI)~\cite{chen2012maximum} heuristic is leveraged to recover a one-bit solution. For the phase shifts design, an accelerated projected gradient (APG) method is applied. Since both MD and APG are first-order methods with closed-form updates, the alternating optimization can be performed very efficiently. Simulation results demonstrate the effectiveness of our proposed joint design and show significant SEP performance gains over conventional linear precoding with  one-bit quantization and the existing one-bit SLP schemes.\vspace{-5pt}

\section{System Model and Problem Formulation} \label{sec:system_model}
Consider a downlink MISO wireless communication system, where an IRS with $N$ reflecting elements is deployed to assist the communications from an $M$-antenna BS to $K$ single-antenna users. The BS makes use of the symbol information and the CSI to perform one-bit precoding for $K$ data streams, one for each user, simultaneously. For simplicity, we assume that perfect CSI is available at the BS\footnote{For the BS-user links, the CSI can be obtained by using conventional estimation method, e.g.,~\cite{lu2014overview}, while for the IRS-related links, CSI can be directly estimated from the concatenated BS-IRS-user links, say by tensor factorization~\cite{de2020parafac}.}. Denote the baseband equivalent channels from BS to user $k$, from BS to IRS, and from IRS to user $k$ by $\bm h_{d, k}\in \mathbb{C}^{M\times 1}$, $\bm G\in \mathbb{C}^{N\times M}$, and $\bm h_{r, k}\in \mathbb{C}^{N\times 1}$, $\forall k\in \mathcal{K}\triangleq \{1,\dots, K \}$, respectively (resp.).



Let $\bm \theta=[\theta_1, \dots, \theta_N]^T$, $\theta_n\in \mathcal{F} \triangleq \{\theta\in \mathbb{C}| \left| \theta\right|=1 \}$, $\forall~n$. Define $\bm \Theta=\sqrt{a} \text{Diag}(\bm \theta)$, where $\text{Diag}(\bm \theta)$ denotes a diagonal matrix with diagonal entries $\bm \theta$, as the phase shift matrix of IRS with
$\theta_n$ and $a \in [0,1]$ being the phase shift and reflecting amplitude of each IRS element, resp. For simplicity, we consider constant reflecting amplitude $a = 1$ and focus on designing $\bm \theta$. Assuming frequency-flat fading channels, the received signal at user $k$ at symbol time $t$ is given by
\begin{equation}\label{eq:receive_signal1}
	\begin{aligned}
	y_{k, t} & =\bm h_{d, k}^H \bm x_t + {\bm h_{r, k}^H \bm \Theta \bm G} \bm x_t+n_{k, t}  \\
	& = (\bm h_{d, k}^H +\bm \theta^T \bm W_{r, k}^H \bm G) \bm x_t +n_{k, t},
	\end{aligned}
\end{equation}
for $k \in \mathcal{K} ,~ t\in \mathcal{T}\triangleq \{1,\dots ,T\}$, where $\bm W_{r, k} = \text{Diag}(\bm h_{r, k})$ and $\bm x_t$ is the transmitted signal at symbol time $t$; $n_{k,t}\sim \mathcal{CN}(0, \sigma^2 )$ is the additive white Gaussian noise (AWGN); $T$ is the length of the transmission block. Under the one-bit signaling, the transmitted signal $\bm x_t$ takes the following form:
\begin{equation*}
	\bm x_t \in \mathcal{X}^M,~ \mathcal{X} \triangleq \{\pm \sqrt{P/2M} \pm \frak{j} \sqrt{P/2M}\},
\end{equation*}
for $t=1,\ldots,T$, where $P$ is the total transmit power.

The idea of one-bit SLP is to judiciously choose $\{\bm x_t\}_{t=1}^T$ and $\bm \theta$ so that user $k$'s received signal $y_{k,t}$ is close to the desired symbol $s_{k,t}~\forall~k\in {\cal K}, t\in {\cal T}$.
In this paper, we consider the $L$-ary PSK constellation, i.e., $s_{k, t}\in \mathcal{S}\triangleq \{s | s = e^{\frak{j}\ell (2 \pi / L)},~\ell=0, \dots, L-1\}$,
where $L$ is the number of constellation points. At the receiver, each user detects their own symbol stream via $\hat{s}_{k, t} = \text{dec}(y_{k, t})$, where $\text{dec}: \mathbb{C} \rightarrow \mathcal{S}$ denotes the $L$-ary PSK decision function, i.e., if the phase angle of $y_{k,t}\in [ \frac{2 \pi}{L} \hat{\ell} -\frac{\pi}{L}, \frac{2 \pi}{L} \hat{\ell} + \frac{\pi}{L}]$, then $\hat{s}_{k, t} = e^{\frak{j} \frac{2\pi}{L} \hat{\ell}}$. Let
\begin{equation} \label{eq:SEP}
	\text{SEP}_{k, t} = \text{Pr}(\hat{s}_{k, t} \neq s_{k, t} | s_{k, t} )
\end{equation}
be the SEP conditioned on $s_{k, t}$. The problem of one-bit SLP with IRS is formulated as
\begin{subequations}\label{eq:SEPmini_problem}
	\begin{align}
		\min_{\bm \theta,\bm X}&~ \max_{k\in \mathcal{K},t\in \mathcal{T}} ~~ \text{SEP}_{k, t} \label{eq:SEPmini_problem_a} \\
        &~~ {\rm s.t.}~~~ ~\bm x_t \in \mathcal{X}^M, ~ t = 1,\cdots, T, \label{eq:SEPmini_problem_b} \\
        & \hspace{35pt}  \left|\theta_n \right| =1,~ n=1,\cdots, N, \label{eq:SEPmini_problem_c}
	\end{align}
\end{subequations}
where $\bm X = [\bm x_1,\cdots, \bm x_T]$. Problem~\eqref{eq:SEPmini_problem} aims to minimize the worst SEP among all the users during the whole transmission block, while satisfying the one-bit and phase-shift constraints.

One of the challenges of problem~\eqref{eq:SEPmini_problem} is that the SEP~\eqref{eq:SEP} for $L$-ary PSK symbols does not admit a simple expression. Instead of tackling problem~\eqref{eq:SEPmini_problem} directly, we first convert  $\text{SEP}_{k, t}$ into a more  tractable form by applying the following upper bound approximation~\cite{shao2018multiuser}:
\begin{equation}
	\text{SEP}_{k, t} \leq 2 Q\left(\frac{\alpha_{k, t}}{\sigma / \sqrt{2}} \sin \frac{\pi}{L} \right),
\end{equation}
where $Q(x) = \int_x^{\infty} \frac{1}{\sqrt{2 \pi}} e^{-z^2 / 2} dz$ and
\begin{equation} \label{eq:alpha1}
	\begin{aligned}
	\alpha_{k, t} & = \mathfrak{R}\left\{(\bm h_{d, k}^H +\bm \theta^T \bm W_{r, k}^H \bm G) \bm x_t s_{k, t}^* \right\} \\  & \hspace{15pt} - \left|\ \mathfrak{I}\left\{(\bm h_{d, k}^H +\bm \theta^T \bm W_{r, k}^H \bm G) \bm x_t s_{k, t}^* \right\}\right| \cot (\pi / L).		
	\end{aligned}
\end{equation}

Since $Q(\cdot)$ is monotonically decreasing, problem~\eqref{eq:SEPmini_problem} can be approximated as
\begin{subequations}\label{eq:solvable_problem}
	\begin{align}
		\min_{\bm \theta, \bm X}&~ \max_{k\in \mathcal{K},t\in \mathcal{T}} ~~ - \alpha_{k, t}\label{eq:solvable_problem_a} \\
        {\rm s.t.} & ~~~ ~\eqref{eq:SEPmini_problem_b}~\text{and}~\eqref{eq:SEPmini_problem_c}.
	\end{align}
\end{subequations}

In the next section, we will focus on solving problem~\eqref{eq:solvable_problem}.


\section{An Alternating Optimization Approach to \eqref{eq:solvable_problem}} \label{sec:approach}

Problem~\eqref{eq:solvable_problem} is essentially a  mixed integer nonlinear program (MINLP), which is generally NP-hard. However, since the variables  are decoupled in the constraints,
we apply the alternating optimization (AO) method to alternately optimize $\bm X$ and $\bm \theta$.

\subsection{Optimizing $\bm X$ with fixed $\bm \theta$} \label{sec:x_design}
Define $\bm h_k^H = \bm h_{d, k}^H +\bm \theta^T \bm W_{r, k}^H \bm G$ and $\bar{\bm x}_t = \left[\mathfrak{R}(\bm x_t)^T, \mathfrak{I}(\bm x_t)^T \right]^T$. For fixed  $\bm \theta$, optimizing problem~\eqref{eq:solvable_problem} with respect to (w.r.t.) $\bm X$ amounts to solving the following $T$ independent problems, one for each $\bar{\bm x}_t$:
\begin{subequations}\label{eq:TranSig_design}
	\begin{align}
		\min_{\bar{\bm x}_t}&~~ \{\max_{k=1, \dots, 2K} ~~ \bm c_k^T \bar{\bm x}_t \} \label{eq:TranSig_design_a} \\
       {\rm s.t.} &~~\bar{\bm x}_t \in \mathcal{L}^{2M}, \label{eq:TranSig_design_b}
	\end{align}
\end{subequations}
for $t= 1,\cdots,T$, where $\mathcal{L} \triangleq \{+ \sqrt{{P}/{2 M}}, ~-\sqrt{{P}/{2 M}} \} ,$
\begin{align*}
	\bm a_k & = \left[ \mathfrak{R}(s_{k, t}^* \bm h_k^H), - \mathfrak{I}(s_{k, t}^* \bm h_k^H)\right]^T,~ \forall k \in \mathcal{K},~ \forall t \in \mathcal{T}, \\
	\bm b_k & = \cot(\pi / L) \left[ \mathfrak{I}(s_{k, t}^* \bm h_k^H), \mathfrak{R}(s_{k, t}^* \bm h_k^H)\right]^T, \forall k \in \mathcal{K}, ~ \forall t \in \mathcal{T}, \\
	\bm c_k & = \left \{\begin{array}{l}
  - \bm a_k + \bm b_k,~~ k=1, \dots, K, \\
  - \bm a_{k-K} - \bm b_{k-K},~~ k=K+1, \dots, 2 K.
\end{array} \right.
\end{align*}
For notational brevity, we will suppress the subscript ``$t$'' in the following derivation.

Problem~\eqref{eq:TranSig_design} is a binary problem, which can be handled by the gradient extrapolated majorization-minimization (GEMM) method in~\cite{shao2019framework}. Here, by exploiting the problem structure, we propose another algorithm for problem~\eqref{eq:TranSig_design} from the dual optimization perspective. Let us first relax the one-bit constraint and consider the following problem:
\begin{equation} \label{eq:regularized_form}
	\min_{\bar{\bm x} \in \tilde{\mathcal{L}}^{2M}}~~ \{\max_{k=1, \dots, 2K} ~~ \bm c_k^T \bar{\bm x}\} + \frac{\mu}{2} \|\bar{\bm x}\|^2
\end{equation}
where $\tilde{\mathcal{L}} = [-\sqrt{P/2M}, \sqrt{P/2M} ]$ is convex relaxation of $\mathcal{L}$; the regularization term $({\mu}/{2}) \|\bar{\bm x}\|^2$ with $\mu>0$ is added to make  problem~\eqref{eq:regularized_form} strongly convex so that its dual problem is smooth. To proceed, we draw a key observation:
\begin{equation} \label{eq:dual_form}
	\max_{k=1, \dots, 2K} ~~ \bm c_k^T \bar{\bm x} = \max_{\bm \lambda \geq \bm 0, \bm \lambda^T\bm 1 = 1} \bm \lambda^T \bm C^T \bar{\bm x},
\end{equation}
where $\bm C = [\bm c_1,\cdots, \bm c_{2K}]$ and $\bm \lambda = [\lambda_1,\cdots,\lambda_{2K} ]^T$. By substituting~\eqref{eq:dual_form} into problem~\eqref{eq:regularized_form}, we have
\begin{subequations}
	\begin{align}
		\eqref{eq:regularized_form} & \Longleftrightarrow \min_{\bar{\bm x}\in \tilde{\mathcal{L}}^{2M}}~ \max_{\bm \lambda \geq \bm 0, \bm \lambda^T\bm 1 = 1} ~ \bm \lambda^T \bm C^T \bar{\bm x} + \frac{\mu}{2} \|\bar{\bm x}\|^2 \\
		& \Longleftrightarrow \max_{\bm \lambda \geq \bm 0, \bm \lambda^T\bm 1 = 1}~ \min_{\bar{\bm x}\in \tilde{\mathcal{L}}^{2M}} ~ \bm \lambda^T \bm C^T \bar{\bm x} + \frac{\mu}{2} \|\bar{\bm x}\|^2, \label{eq:TranSig_design_dual}	
\end{align}
\end{subequations}
where~\eqref{eq:TranSig_design_dual} is due to Sion's minimax theorem~\cite{sion1958general}.
 One can easily show that the optimal solution for the inner minimization in~\eqref{eq:TranSig_design_dual} is given by
 \begin{equation} \label{eq:x_bar}
 	\bar{\bm x}^{\star} = - \big[(1/\mu) \bm C \bm \lambda \big]_{-\bm {\sqrt{P/2M}}}^{\bm {\sqrt{P/2M}}},
 \end{equation}
 where $[\cdot]_{\bm a}^{\bm b}$ is the elementwise clipping function and $[x_i]_a^b = \text{max}\{a,\text{min}\{x_i, b\}\}$. By substituting~\eqref{eq:x_bar} into~\eqref{eq:TranSig_design_dual}, the dual problem can be simplified as
 \begin{equation}\label{eq:lambda_design}
 			\min_{\bm \lambda \geq \bm 0, \bm \lambda^T\bm 1 = 1}~~ f_{\mu}(\bm \lambda) \triangleq \sqrt{\frac{P}{2M}}  \sum_{m=1}^{2M} h_{\mu \sqrt{\frac{P}{2M}} }\left(\bar{\bm c}_m \bm \lambda \right),
 \end{equation}
 where $\bar{\bm c}_m$ represents the $m$-th row of $\bm C$; $h_{\rho}$ is the (smooth) Huber function defined as
\begin{equation*}
	 h_{\rho}(y) = \left \{\begin{array}{l}
  y^2/2\rho,~~ |y|\leq \rho, \\
  |y|-\rho/2,~~ \text{otherwise}. \end{array} \right. 	
\end{equation*}
We should mention that the dual problem~\eqref{eq:lambda_design} has $2K$ variables, which is usually much smaller than $2M$ in~\eqref{eq:regularized_form}, especially for the massive MIMO regime.


Notice that problem~\eqref{eq:lambda_design} has a smooth objective and a probability simplex constraint. We apply the mirror descent (MD) method~\cite{beck2003mirror}  to find an optimal solution for it. The MD method is also known as generalized projected gradient method (PGM). By introducing a more general distance measure other than Euclidean distance, the MD method can better exploit the probability simplex structure so that the projection can be computed in closed form, thereby achieving more efficient updating than PGM. Readers are referred to~\cite{beck2003mirror} for more details on MD. Herein we only present the key steps of the MD method. The MD method solves problem~\eqref{eq:lambda_design} by iteratively performing the following updates:
\begin{equation} \label{eq:MD_update}
	\bm \lambda^{r+1} = \arg \min_{\bm \lambda \geq \bm 0, \bm \lambda^T\bm 1 = 1} \big<\nabla f_{\mu}(\bm \lambda^r), \bm \lambda \big> + \frac{1}{2 \eta^r} B(\bm \lambda,\bm \lambda^r)
\end{equation}
for $r= 0,1,2,\cdots$ until convergence, where $\eta^r$ is the step size and can be determined by backtracking line search~\cite{beck2009fast}; $B(\bm \lambda,\bm \lambda^r)$ is the distance measure, which is chosen as $\sum_{k=1}^{2K} \lambda_k \log \left(\lambda_k/\lambda_k^r \right) $, i.e., the Kullback-Leibler (KL) divergence, such that~\eqref{eq:MD_update} can be expressed in closed form~\cite{beck2003mirror}:
\begin{equation} \label{eq:lambda_update}
	\lambda_k^{r+1} = \frac{\lambda_k^{r} \exp (-\eta^r \nabla_{\lambda_k} f_{\mu} (\bm \lambda^{r}))}{\sum_{j=1}^{2K}\lambda_j^r \exp (-\eta^r \nabla_{\lambda_j} f_{\mu} (\bm \lambda^{r}) ) },k=1,\cdots, 2K,		
\end{equation}
where $\nabla_{\lambda_k} f_{\mu} (\bm \lambda) = \frac{\partial f_{\mu} (\bm \lambda)}{\partial \lambda_k}$. Once we obtain the optimal solution $\bm \lambda^{\star}$, the optimal $\bar{\bm x}^{\star}$ for problem~\eqref{eq:regularized_form} can be recovered via~\eqref{eq:x_bar}.

Notice that $\bar{\bm x}^\star$ is a solution of the relaxed one-bit SLP problem. It is possible that some of the elements of  $\bar{\bm x}^\star$ do not comply with the one-bit constraints. Fortunately, it is recently shown in~\cite{li2020near} that the number of elements violating the one-bit constraints are no greater than $2K-1$. That is, for massive MIMO with small number of users we can recover most of the one-bit elements from $\bar{\bm x}^\star$. In the following, we propose a heuristic, yet efficient, approach to project the remaining elements onto the one-bit set. Denote $\mathcal{I} \triangleq\{m ~|~ |\bar{x}_m^\star|<\sqrt{{P}/{2M}},m=1,\ldots, 2M \}$ and $\varphi(\bar{\bm x}_{\cal I}, \bar{\bm x}_{-\cal I}) \triangleq \max_{k=1,\ldots, 2K} \bm c_k^T \bar{\bm x}$, i.e., the objective in~\eqref{eq:TranSig_design_a}. By fixing $\bar{\bm x}_{-\cal I} = \bar{\bm x}^\star_{-\cal I}$, we consider the following problem:
	\begin{equation} \label{eq:mbi}
	 \min_{\bar{\bm x}_{\cal I} \in {\cal L}^{|\cal I|}} ~ \varphi(\bar{\bm x}_{\cal I}, \bar{\bm x}_{-\cal I}^\star)
	 \end{equation}
We employ the MBI method to approximately solve the above problem. The idea of MBI is similar to block coordinate descent (BCD) method, except that for every updating cycle, only the coordinate that leads to the maximum improvement of the objective value is updated and the remaining coordinates are kept unchanged. MBI is shown to guarantee better convergence than BCD; readers are refer to~\cite{chen2012maximum} for more details on MBI. As for  problem~\eqref{eq:mbi}, since each coordinate has only two possible choices, the MBI method can work very efficiently by cyclically performing binary tests for each coordinate. Moreover, it is also helpful to run multiple initializations of MBI to avoid bad local minimum. In our simulations, we found that five times random initializations are enough to output a satisfactory one-bit solution.

\subsection{Optimizing $\bm \theta$ with fixed $\bm X$} \label{sec:approach_B}
 Define $\bm u_{k, t} = \bm W_{r, k}^H \bm G \bm x_t s_{k, t}^*$ and $v_{k, t} = \bm h_{d, k}^H \bm x_t s_{k, t}^*$. The $\alpha_{k, t}$ in~\eqref{eq:alpha1} can be rewritten as
\begin{equation*}
	\alpha_{k, t} = \mathfrak{R}\{\bm \theta^T \bm u_{k, t} + v_{k, t} \} - \left| \mathfrak{I}\{\bm \theta^T \bm u_{k, t} + v_{k, t} \} \right| \cot (\pi/ L),
\end{equation*}
for all $k$ and $t$. Let $\bm{\bar{\theta}} = [\mathfrak{R}(\bm \theta )^T, \mathfrak{I}(\bm \theta)^T ]^T$ and through complex-to-real transformation, the phase-shift design problem (with fixed $\bm X$) is given by
\begin{subequations}\label{eq:phaseShift_pro}
	\begin{align}
		& \min_{ \bm{\bar{\theta}} \in \mathbb{R}^{2N}}~\big\{ \max_{\substack{k=1, \dots, 2 K,\\ t=1, \dots, T, }} ~~\bm{\bar{\theta}}^T \bm \eta_{k, t} + \bar{v}_{k, t}\big\} \label{eq:phaseShift_pro_a} \\
        &  {\rm s.t.}~ \bar{\bm \theta} \in \bm \Phi \triangleq \left\{\bar{\bm \theta}~| ~|\bar{\theta}_n |^2 + |\bar{\theta}_{n+N} |^2 = 1, ~\forall n  \right\}, \label{eq:phaseShift_pro_b}
    \end{align}
\end{subequations}
where
\begin{align*}
	\bm \eta_{k, t} & = \left \{\begin{array}{l}
  - \bm q_{k, t} + \bm p_{k, t},~~ k=1, \dots, K, \\
  - \bm q_{k-K, t} - \bm p_{k-K, t},~~ k=K+1, \dots, 2 K. \end{array} \right. , \forall t \\
  	\bm q_{k, t} & = \left[ \mathfrak{R}(\bm u_{k, t})^T, - \mathfrak{I}(\bm u_{k, t})^T\right]^T, ~ \forall k \in \mathcal{K} ,~\forall t \in \mathcal{T} \\
	\bm p_{k, t} & = \cot(\pi / L) \left[ \mathfrak{I}(\bm u_{k, t})^T, \mathfrak{R}(\bm u_{k, t})^T\right]^T,~ \forall k \in \mathcal{K} ,~\forall t \in \mathcal{T} \\
	\bar{v}_{k, t} &\! = \! \left \{\begin{array}{l}
 \hspace{-8pt} - \mathfrak{R}(v_{k, t}) + \mathfrak{I}(v_{k, t})\cot (\pi / L),~~ k=1, \dots, K, \\
 \hspace{-8pt} - \mathfrak{R}(v_{k-K, t}) \! - \! \mathfrak{I}(v_{k-K, t})\cot (\pi / L), k \! = \! K \! + \! 1, \dots, 2 K. \end{array} \right.
\end{align*}

Since  $\bm \Phi$ is a unit modulus set,  projection onto  $\bm \Phi$ can be easily computed by normalization. This motivates us to apply the APG method to handle problem~\eqref{eq:phaseShift_pro}. Specifically,  we first replace~\eqref{eq:phaseShift_pro_a} with its smooth approximation:
\begin{equation*}
	h(\bar{\bm \theta}) = \delta \log \left(\sum_{t=1}^T \sum_{k=1}^{2K} \exp \left(\frac{\bm{\bar{\theta}}^T \bm \eta_{k, t} + \bar{v}_{k, t}}{\delta} \right) \right),
\end{equation*}
where $\delta > 0$ is the smoothing parameter controlling the accuracy of the approximation. In particular, the approximation is accurate as $\delta \rightarrow 0$. Then, the APG update for $\bar{\bm \theta} $ is performed:
\begin{equation} \label{eq:theta_APG}
		\bar{\bm{\theta}}^{r+1}= \Pi_{\bm \Phi} \left(\bm z^r - \frac{1}{\tau^r} \nabla_{\bar{\bm{\theta}}} h(\bm z^r)\right),~~ r=0,1,2, \cdots,
\end{equation}
until convergence, where $\bm z^r$ is the extrapolated point
\begin{equation*}
	\bm z^r = \bar{\bm{\theta}}^{r} + \psi_r ( \bar{\bm{\theta}}^{r} -  \bar{\bm{\theta}}^{r-1});
\end{equation*}
with
\begin{equation} \label{eq:psi}
\psi_r = \frac{\zeta_r-1}{\zeta_r},~\zeta_r = \frac{1+\sqrt{1+4 \zeta_{r-1}^2}}{2},~ \zeta_{-1}=0,
\end{equation}
and $1/\tau^r$ is the step size; $\Pi_{\bm \Phi}$ is elementwise projection operator on $\bm \Phi$, which can be obtained via normalization.

\section{Numerical Results} \label{sec:num_results}
In this section, we evaluate the performance of our proposed algorithm by simulations. The simulation settings are as follows: the locations of the BS and the IRS are set to $(0,0)$ and $(20,10)$, resp., and users are randomly distributed in a circle area centered at $(30,0)$ with radius 10m, i.e., the IRS is deployed closer the users to create rich scattering propagation environment. We set $K=14$, $M=128$, $T = 100$ and $P = 20\text{dB}$. The path loss of the large-scale fading is modeled as $ L(d) = C_0 \varrho d^{-\varsigma}$, where $C_0$ denotes the pass loss at the reference distance 1 meter, $\varrho$ denotes the product of the source and the terminal gain, $d$ and $\varsigma$ represent the link distance and the path loss exponent, resp. In particular, for the BS-user link we set $C_0 \varrho_{B,k}= -15\text{dB}$ and $\varsigma_{B,k} = 3.2$, $\forall k\in \mathcal{K}$, while for the IRS-aided link, the concatenate path loss is set as  $C_0^2 \varrho_{B,I} \varrho_{I,k} (d_{B,I})^{-\varsigma_{B,I}} (d_{I,k})^{-\varsigma_{I,k}}$ with $C_0^2 \varrho_{B,I} \varrho_{I,k} = -20\text{dB}$ and $\varsigma_{B,I} = \varsigma_{I,k} = 2.2$. To characterize small-scale fading, we assume Rayleigh fading model for all channels. The regularization parameter and the smoothing parameter are chosen as $\mu =5\times 10^{-4}$ and $\delta= 10^{-2}$, resp..


 We compare our proposed design, which is labeled ``1-bit SLP with IRS, MD'' with the following benchmark schemes:
 1) ``1-bit SLP with IRS, GEMM", which alternately optimizes the one-bit signal by the GEMM method in~\cite{shao2019framework} and the phase shifts  by the APG method. We should mention that while this scheme is straightforward, it is  new and has not been reported in the current literature; 2) ``SLP with IRS (infinite bits)'', which was proposed in~\cite{shao2020minimum} by assuming that the precoder has infinite resolution. Since ``SLP with IRS (infinite bits)" relaxes the one-bit constraints on $\bm X$, it serves as the performance lower bound; 3)  ``SLP with IRS (1-bit quant.)'', which first runs the design in~\cite{shao2020minimum}, followed by naive one-bit quantization on the precoder; 4) ``ZF with IRS (1-bit quant.)", which adopts conventional zero-forcing (ZF) linear precoder, followed by one-bit quantization; 5) ``1-bit SLP w/o IRS, GEMM''; 6) ``SLP w/o IRS (1-bit quant.)'' and 7)  ``1-bit SLP w/o IRS, MD''. Schemes 5)-7) are the counterparts of schemes 1), 3) and the proposed ``1-bit SLP with IRS, MD'' in the absence of IRS, resp.
All the results were obtained by averaging over 1,000 channel realizations. For designs with IRS, the outer AO loop terminates when $\|\bm X^i-\bm X^{i-1}\|^2+\|\bm \theta^{i} - \bm \theta^{i-1} \|^2 < 10^{-4}$ or  the number of iterations exceeds 20.

\begin{figure}[!h]
	\centerline{\resizebox{.49\textwidth}{!}{\includegraphics{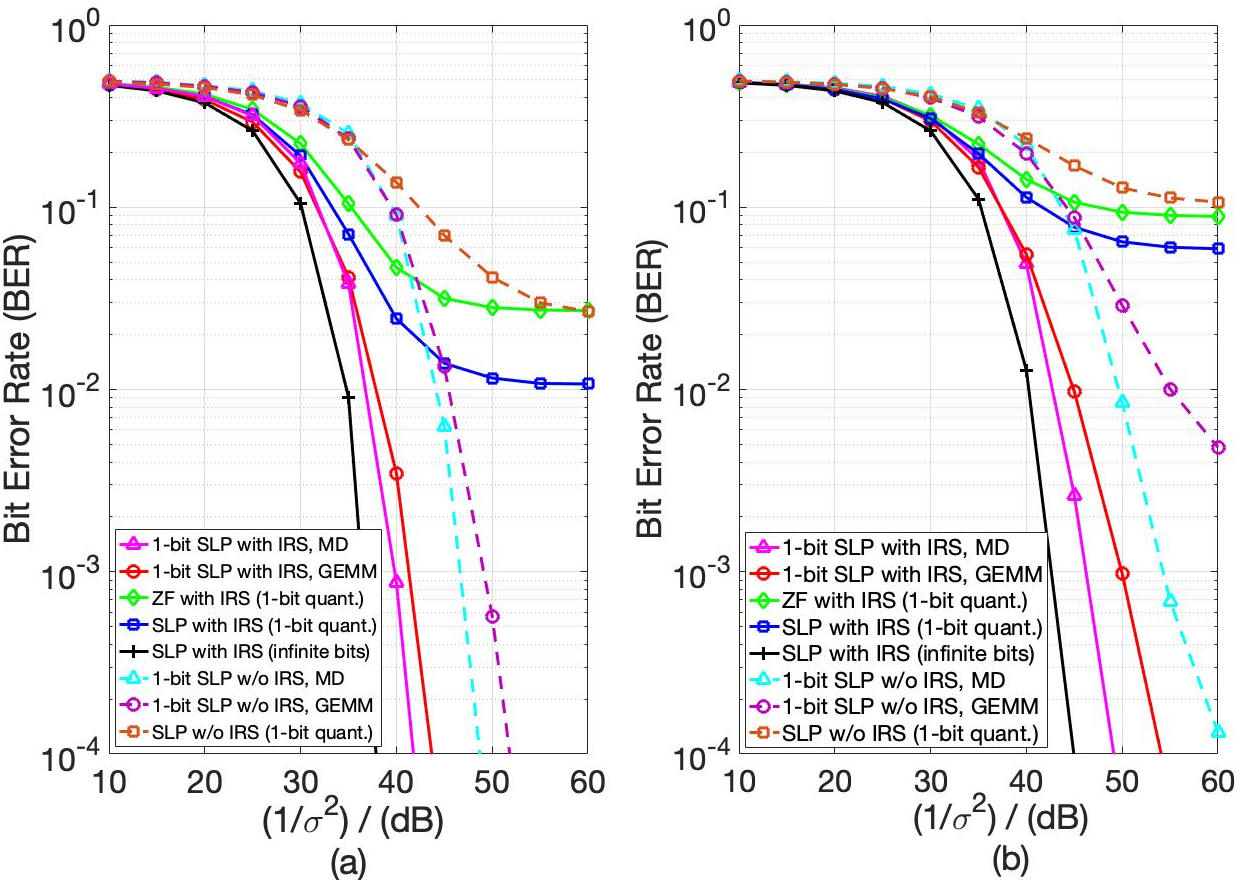}}
	}   \caption{BER performance vs. $1 / \sigma^2$; (a) QPSK; (b) 8-ary PSK.} \label{fig:QPSK_8PSK}
	\vspace*{-1.4\baselineskip}
\end{figure}

 Fig.~\ref{fig:QPSK_8PSK}(a) and Fig.~\ref{fig:QPSK_8PSK}(b) show the bit-error rate (BER) performance versus the reciprocal of the noise power $1 / \sigma^2$ for QPSK and 8-ary PSK, resp., when fixing the number of IRS elements to $N = 32$. First, it can be observed that both our proposed algorithm and ``1-bit SLP with IRS, GEMM'' design outperform ``SLP with IRS (1-bit quant.)'' and ``ZF with IRS (1-bit quant.)''. This reveals that it is crucial to explicitly take into account the one-bit constraints during the optimization process, and naive one-bit quantization could incur large performance loss. Also, our proposed design achieves even better BER performance than ``1-bit SLP with IRS, GEMM" at high SNR region. Moreover, our proposed joint one-bit SLP and IRS design works much better than 1-bit SLP without IRS designs (i.e., ``1-bit SLP w/o IRS, MD'' and ``1-bit SLP w/o IRS, GEMM''), which validates the efficacy of IRS in improving the system performance.

Table~\ref{tab:1} compares the average running times of the proposed design and ``1-bit SLP with IRS, GEMM"  under the same setting in Fig.~\ref{fig:QPSK_8PSK}. It is seen that the proposed design is faster than the GEMM-based one.

\begin{table}[!h]
\centering
\caption{Average runtime (Sec.), $(M,N,K)=(128,32,14)$}\label{tab:1}
\begin{tabular}{|p{3.5cm}|p{2cm}<{\centering}|p{2cm}<{\centering}|}
  \hline
  \multirow{2}{*}{Algorithm} & \multicolumn{2}{c|}{Modulation}  \\ \cline{2-3}
         & QPSK & 8PSK \\ \hline
  1-bit SLP with IRS, MD &     1.5115 &   1.7317   \\
  1-bit SLP with IRS, GEMM &     2.4692   &   2.5511 \\
  \hline
\end{tabular}
\vspace*{-1\baselineskip}
\end{table}

\section{Conclusions} \label{sec:conclusions}
This paper considers a joint one-bit SLP and phase shifts design for the IRS-aided MU-MISO downlink communications. A minimum-SEP-based formulation is studied for PSK constellation. An alternating optimization approach is proposed to alternately optimize the one-bit signal  and the phase shifts.
Simulation results demonstrate the efficacy of incorporating IRS to improve the system BER performance. Moreover, by explicitly taking into account the symbol information and the one-bit constraints, the proposed design attains superior performance as compared with conventional
linear precoder and SLP with naive one-bit quantization.


\bibliography{maintext}

\end{document}